\documentclass[letterpaper,superscriptaddress,aps,prl,nolongbibliography,twocolumn,showpacs,floatfix]{revtex4-1} 

\usepackage{graphicx,helvet}
\usepackage{color}
\usepackage{bm}                      
\usepackage{tikz}
\usepackage{amsfonts}
\usepackage{lipsum}
\definecolor{mygray}{gray}{0.4}
\definecolor{light-blue}{rgb}{0.8,0.85,1}
\usepackage{amsmath}%
\usepackage{MnSymbol}%
\usepackage{wasysym}%
 \usepackage{soul}

\usetikzlibrary{decorations.shapes}

\newcommand{\beqa}{\begin{eqnarray}}
\newcommand{\eeqa}{\end{eqnarray}}
\newcommand{\beq}{\begin{equation}}
\newcommand{\eeq}{\end{equation}}
\newcommand{\ktimes}{\rangle\! \langle}
\newcommand{\op}[2]{|#1\ktimes #2|}
\usepackage{amsmath}
\DeclareMathOperator{\sign}{sign}

\newcommand{\tr}{\mathop{\mathrm{Tr}}\nolimits}

\newcommand{\ket}[1]{{\vert #1 \rangle}}
\newcommand{\bra}[1]{{\langle #1 \vert}}

\newcommand{\ifimar}{Instituto de Investigaciones F\'isicas de Mar del Plata (IFIMAR), CONICET-UNMdP,  Mar del Plata,  Argentina}
\newcommand{\conicet}{Consejo Nacional de Investigaciones Cient\'ificas y Tecnol\'ogicas (CONICET), Argentina}
\newcommand{\uba}{Departamento de F\'isica ``J. J. Giambiagi'' and IFIBA, FCEyN, Universidad de Buenos Aires, 1428 Buenos Aires, Argentina}
\begin{document}
\title{Quantum to classical transition in the work distribution for chaotic systems}
\author{Ignacio Garc\'ia-Mata} \affiliation{\ifimar} \affiliation{\conicet}
\author{Augusto J. Roncaglia}\affiliation{\uba}
\author{Diego A. Wisniacki} \affiliation{\uba}

\begin{abstract}
The work distribution is a fundamental quantity in nonequilibrium 
thermodynamics mainly due to its connection with fluctuations theorems.
Here we develop a semiclassical approximation to the work distribution
for a quench process in chaotic systems. The approach is based on the dephasing
representation of the quantum Loschmidt echo and on the quantum ergodic conjecture, which states that the Wigner function
of a typical eigenstate of a classically chaotic Hamiltonian is equidistributed on the energy shell.
We show that our semiclassical approximation is accurate in describing the quantum distribution as we increase the temperature.   
Moreover, we also show that this semiclassical approximation provides a link between the quantum and classical work distributions.
 \end{abstract}
%
%
\maketitle
During the last years there was an increasing interest in the field of thermodynamics and statistical physics \cite{Campisi2011,Goold16}. 
This rebirth of the area  has been fuelled by the technological advances that lead to control with 
extreme precision the dynamics of small quantum systems.
In the context of nonequilibrium statistical physics, standard concepts such as work and heat work are random variables characterised by a distribution, 
and have been redefined so as to include quantum effects \cite{Kurchan2000,Tasaki00,Talkner07}.
Notably, fluctuation theorems such as Jarzynski  \cite{Jarzynski1997} and Crooks \cite{Crooks1999} equalities, have been 
extended to the quantum regime. 
Despite the similarities between standard and quantum fluctuation theorems, the transition between quantum and 
classical descriptions seems to be elusive, and this difference resides in the quantum definition of work. 

The accepted definition of quantum work, performed or extracted after a single realization of a 
process in an isolated system, can be formulated in terms of two projective  measurements of the system's energy
or two-point measurement \cite{Kurchan2000,Tasaki00,Talkner07}.
In this way, the fluctuations in the value of work have thermal and quantum  origins. 
However, it could also be argued that this definition is arbitrarily proposed so that the fluctuation
relations are fulfilled. It has been thus suggested \cite{Jarzynski2015,Quan2016} 
that a justification of this definition would be to test the correspondence
principle in the classical limit. Several studies have considered the quantum and classical distributions of work  in harmonic systems \cite{Deffner08,Talkner08,Campisi08,Deffner10,Ford12,Talkner13}, where analytical solutions are available. 
Interestingly in \cite{Jarzynski2015}, by employing a semiclassical method, it has been shown that 
in one dimensional integrable systems (quartic oscillator) there is a correspondence between classical and quantum transition probabilities.  
On the other hand for chaotic systems, there is numerical evidence showing that  the correspondence principle also applies \cite{Quan2016} supporting the 
quantum definition of work.

Here we go a step further: we 
present a semiclassical expression for the distribution of work
done on a system after a quench for fully chaotic system.
Moreover, we show analytically that the correspondence between the classical an quantum distributions is recovered from the semiclassical expression 
in the limit of vanishing Planck constant. Furthermore we verify   
the good agreement between the semiclassical and quantum 
work distributions, and the quantum-classical correspondence, 
using numerical simulations of a quantum particle inside a  
stadium billiard that suddenly changes its inner potential.  
The main ingredients we used to derive the semiclassical expression are: the connection of the characteristic function with the 
fidelity amplitude  or Loschmidt echo \cite{Silva2008}, the semiclassical dephasing representation \cite{vanicek2003,vanicek2004,vanicek2006} of the  fidelity amplitude
and the conjecture by Berry and Voros that the Wigner functions for eigenstates of chaotic systems  are peaked on the corresponding energy shell \cite{Berry_1977, Voros}.

We consider the following process applied to a quantum system described by a  Hamiltonian $H_{\xi}$ that depends on a control parameter $\xi$.
We assume that  the system is in thermal equilibrium with a bath at inverse  temperature proportional to $\beta^{-1}$ (in the following we will consider $k_B=1$).
The initial state of the system is then $\rho_\beta=\exp(-\beta H_{\xi_0})/Z^Q_{\xi_0}$ and $Z^Q_{\xi_0}=\tr[ \exp(-\beta\,H_{\xi_0})]$.
Next, the system is decoupled from the bath, and  the control parameter is 
suddenly changed from $\xi_0$ to $\xi_f$, taking the system away from equilibrium. 
As a result, the work performed on the system after the quench, is a random quantity given 
by the difference of the outcomes of two energy measurements  $W=E^n_{\xi_f}-E^m_{\xi_0}$, one at the beginning and the other at the end of the process. 
In this way, work is characterized by the following distribution  \cite{Kurchan2000,Tasaki00,Talkner07}:
\begin{equation}
P^Q(W)=\sum_{n,m}P^Q(m) P^Q(n| m)\delta[W-(E^n_{\xi_f}-E^m_{\xi_0})],
\end{equation}
where $P^Q(m)={e^{-\beta E^m_{\xi_0}}}/{Z^Q_{\xi_0}}$, $P^Q(n| m)$ is the quantum conditional probability to obtain 
the eigenstate of $E^n_{\xi_f}$ at the final measurement, given that the initial result was $E^m_{\xi_0}$ (for a quench this is equal to the squared
overlap between the corresponding eigenstates), and $\delta$ is the Dirac $\delta$ function.
Different strategies that enable the reconstruction of the quantum work distribution have been put forward \cite{Huber2008,Dorner2013,Mazzola2013,Roncaglia2014} 
and also recently measured verifying fluctuations theorems in the quantum regime  \cite{Batalhao2014,An2014}. 

In our approach we consider the characteristic function \cite{Talkner07,Deffner2008},  that is defined as the Fourier transform of the work distribution
\begin{equation}
G(u)=\int dW e^{i u W} P(W).
\end{equation}
Notably, the characteristic function can be directly measured using interferometric techniques, as proposed in \cite{Dorner2013,Mazzola2013}
and implemented experimentally in  \cite{Batalhao2014}.
As it was noticed in \cite{Silva2008} the characteristic function for a quantum system that is subjected to a quench can be expressed as
\begin{equation}
\label{gdet}
G^Q(u)=\langle e^{i u H_{\xi_f}}e^{-i u H_{\xi_0}} \rangle=\tr[e^{i u H_{\xi_f} } e^{-i  u H_{\xi_0}} \rho_\beta],
\end{equation}
where $\rho_\beta$ is a thermal state of $H_{\xi_0}$ at temperature $\beta^{-1}$.
In this expression $u$ is a time-like variable so, if it is replaced by $t/\hbar$, Eq.~\eqref{gdet}
can be regarded as the average amplitude probability over two different time evolutions, or simply an averaged 
Loschmidt echo amplitude \cite{Gorin2006,Jacquod2009,DiegoScholar}. 
Remarkably, in \cite{vanicek2003,vanicek2004,vanicek2006}  a very efficient  semiclassical method  
to compute the LE amplitude -- called dephasing representation (DR) -- was proposed.
It is based on the initial value representation \cite{Miller1970,*Miller2001} and its success is partly due to the validity of 
the shadowing theorem \cite{vanicek2004}. Moreover it circumvents some of the problems of the previous semiclassical 
techniques, like the need of using two pairs of classical trajectories, search for periodic orbits, or other special classical 
trajectories. The DR has been successfully used in various areas such as the study of irreversibility and the Loschmidt echo \cite{vanicek2003,vanicek2004,vanicek2006,garciamaNJP,GRW2016}  
the local density of states in chaotic systems \cite{Diego2010} and molecular dynamics\cite{Zimm2010,*Zimm2010_2}. 
 Bellow, we derive a semiclassical expression for the characteristic function that is based on the use of the DR.

Let us start the derivation by considering a quench, 
where a parameter of the  Hamiltonian $H_{\xi}=H_0+\xi\, V$ is suddenly changed from $\xi_0\rightarrow \xi_f$.
Then, the characteristic function for this process can be expressed in the DR as \cite{vanicek2006}
\begin{equation}
\label{eq:DR1}
G^{DR}(u)=\int {d^{2D} x_0}\, {\cal W}_{\beta} (x_0)\exp [i \Delta S(x_0, u \hbar)/\hbar],
\end{equation}
where $D$ is the number of degrees of freedom, $x~\equiv~(q,p)~\in~\mathbb{R}^{2D}$ is a collective notation, 
 $\Delta S(x_0,t)~\equiv~\int_0^t d\tau \Delta H\left(x(\tau)\right)$ is an action difference, 
 $\Delta H \equiv H_{\xi_f}-H_{\xi_0}$, $x(\tau)$ denotes the phase space  coordinates at time $\tau$ of a trajectory generated by 
 the Hamiltonian $H_{\xi_0}$ with the initial condition $x_0$, and ${\cal W}_\beta (x_0)$ is the Wigner function of the thermal state $\rho_\beta$.
\begin{figure}
\includegraphics[width=0.95\linewidth]{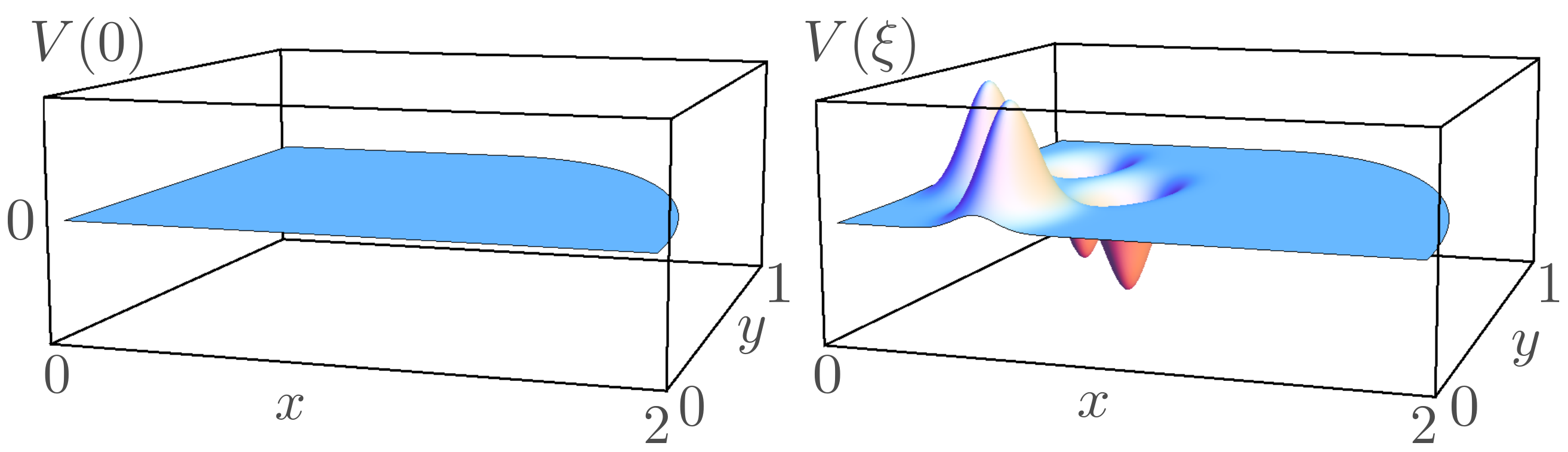}
\caption{Schematic depiction of the quench scheme: $H_0 \rightarrow H_0+V(\xi)$, where $H_0=(p_x^2+p_y^2)$. The Gaussians in the quench potential are centered at
$(x_1,y_1)=(0.2,0.4)$, $(x_2,y_2)=(0.67,0.5)$, $(x_3,y_3)=(0.5,0.15)$ and  $(x_4,y_4)=(0.3,0.75)$. 
\label{fig:stadium}}
\end{figure}
Since the thermal state of the Hamiltonian $H_{\xi_0}$ is $\rho_\beta=~\sum_m e^{-\beta E^m_{\xi_0}}\op{E^m_{\xi_0}}{E^m_{\xi_0}}/Z_{\xi_0}^Q$, 
where $\ket{E^m_{\xi_0}}$ are the eigenstates of $H_{\xi_0}$, its Wigner function can be written as a linear combination of the Wigner functions ${\cal W}_m$ of each eigenstate,
${\cal W}_\beta(x)=  \sum_m {e^{-\beta E^m_{\xi_0}}}/{Z_{\xi_0}^Q}\;{\cal W}_{m}(x)$.
Now comes the key approximation that we use. 
We consider a quantum system with fully chaotic classical counterpart, a system where the quantum ergodic conjecture applies (QEC) \cite{Berry_1977, Voros}.
The QEC states that the Wigner function that represents a typical $E$-energy eigenstate of a classically chaotic Hamiltonian
can be approximated by the classical Liouville probability density for the $E$-energy shell of the Hamiltonian,
\beq
{\cal W}_{m}(x) = \frac{\delta[E^m_{\xi_0}-H_{\xi_0}(x)]}{\int d^{2D}x\; \delta[E^m_{\xi_0}-H_{\xi_0}(x)]}.
\label{wigcha}
\eeq
 In addition, in order to simplify even more the final expression, we will consider that \cite{Jarzynski2015,Quan2016}
 ${e^{-\beta E^m_{\xi_0}}}/{Z^Q_{\xi_0}}~\approx~\int_{E^m_{\xi_0}}^{E^{m+1}_{\xi_0}}{e^{-\beta E}}/{Z^C_{\xi_0}}\, g(E) dE$, 
 where $Z^C_{\xi_0}~=~\int dx\, \exp[-\beta H_{\xi_0}(x)]$ is the classical partition function.
 Applying these approximations to Eq. ~\eqref{eq:DR1} we arrive at the main result of the paper, 
 a semiclassical expression of the characteristic function (see Supplementary material):
\beq
\label{GSC}
G^{SC}(u)=\int {d^{2D} x_0}\, \frac{e^{-\beta H_{\xi_0}(x_0)}}{Z^C_{\xi_0}} \exp[i \Delta S(x_0, u \hbar)/\hbar].
\eeq
It is worth pointing out that the classical work distribution $P^{C}(W)$ \cite{Jarzynski2015,Schwieters95} can be recovered from $G^{SC}(u)$. 
This can be done by taking the limit  $\hbar \rightarrow 0$ and applying the 
inverse Fourier transform to $G^{SC}(u)$ (see Supplementary material):
\beq
P_{\hbar \rightarrow 0}^{SC}(W)= \frac{1}{2 \pi} \int du\; e^{-i u W} \, G_{\hbar \rightarrow 0}^{SC}(u)=P^{C}(W).
\label{PSC}
\eeq
So far, we have derived semiclassical a expression for the characteristic function using several approximations. 
In principle this approach could be applied to fully chaotic systems, bellow we test Eqs. \eqref{GSC} and \eqref{PSC} numerically for a specific model. 

The system that we consider is a quantum particle of mass $m=1/2$ inside a  
desymmetrized stadium billiard with radius $r=1$ and straight line of length $l=1$.  Notice that in this case the hard walls of the billiard do not move in the process.
In Refs. \cite{Quan2012,Quan2016} it was numerically observed that for systems with moving hard walls the  quantum-classical correspondence 
depends strongly on the  adiabaticity of the process. Furthermore, when the boundary is quenched infinitely fast it is shown that there is no connection 
between quantum and classical distributions.
Here we consider a quench that consists of a sudden change of the Hamiltonian parameter, from $\xi_0=0$ to $\xi_f=85$,
with a smooth potential given by four Gaussians: 
$V(\xi)= \xi \sum_{i=1}^4 \sign_i \exp(-[(x-x_i)^2-(y-y_i)^2]/(2 \sigma^2)]$,
$\sigma=0.1$ their widths (see Fig.~\ref{fig:stadium}), and $\sign_i=(-1)^{i-1}$. The Gaussians are centered at
$(x_1,y_1)=(0.2,0.4)$, $(x_2,y_2)=(0.67,0.5)$, $(x_3,y_3)=(0.5,0.15)$ and  $(x_4,y_4)=(0.3,0.75)$. 
\begin{figure}
\includegraphics[width=0.95\linewidth]{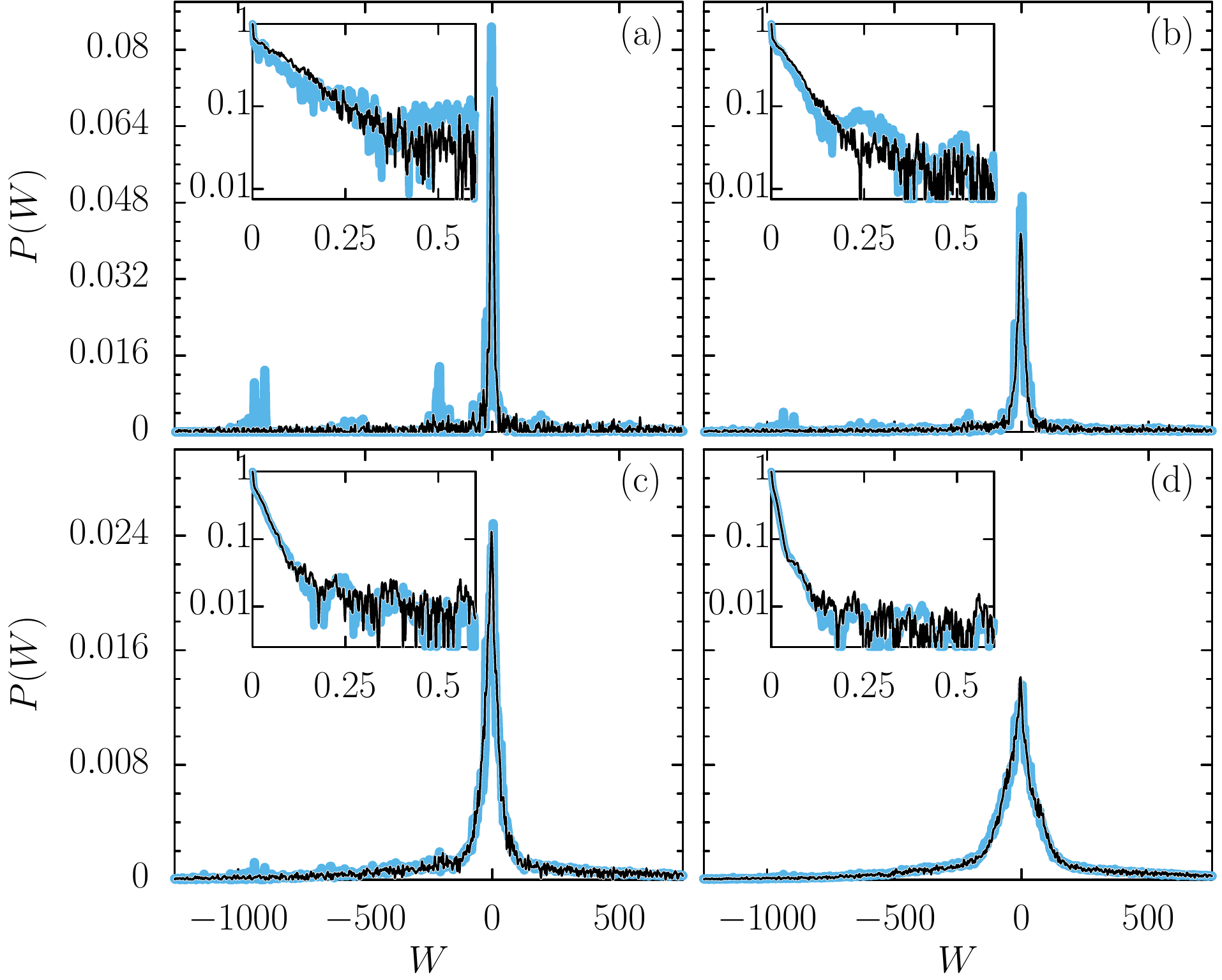}  
\caption{Quantum (light blue/gray) and semiclassical (black) work distribution for different temperatures, from (a)--(d): $\beta=2^{-6},2^{-8},2^{-10}, 2^{-12}$.
In the inset we show the characteristic function for each temperature.
\label{fig:PdeW}}
\end{figure}

In order to obtain the quantum characteristic function $G^Q(u)$ and  $P^Q(W)$, we first compute 
the eigenstates of the unperturbed stadium using the 
scaling method \cite{vergini95}, and then we perform a diagonalization using the eigenbasis of the initial 
Hamiltonian $H_{\xi_0}$ to obtain the quenched system eigenstates. 
We have considered the first 2500 states, but the basis is truncated at 5600 (to ensure the accuracy of the first 2500). 
The semiclassical approximations of  $G^{SC}(u)$ and  $P^{SC}(W)$ are computed using Eq. \eqref{GSC} (and its Fourier Transform)
by sampling over a set of classical trajectories. 
 
In Fig.~\ref{fig:PdeW} we show $P^{Q}(W)$ and $P^{SC}(W)$,
where different values of temperature $\beta^{-1}$  are considered, and $\hbar=1$.  In the inset we show the corresponding characteristic functions. 
The solid light-blue (gray) lines correspond to the quantum result and the black curves the semiclassical one. We remark that the accuracy of 
the semiclassical calculations does not depend very much on the number of trajectories that we use \cite{Zambrano2013}. In particular, the results shown  
in Fig.~\ref{fig:PdeW} were obtained using only $\approx 9\times 10^4$  randomly chosen initial conditions. 
We can see that the main features of the work distribution are well reproduced for all these temperatures. 
In the case of  $\beta=2^{-6}$ we can see that the semiclassical distribution deviates from the quantum one. 
This is due to the fact that the proportion of low-lying energy eigenstates contributing to the distribution is significant, making our approximation less accurate. 
If $\Delta E=(4 \pi/  \hbar^2)/(2 m A)$ is the mean level spacing of a quantum billiard of area $A$ and mass $m$, then for $\beta=2^{-6}$ 
the relevant number of states is $\sim 10$, so only a few,  low-lying energies contribute.
On the other hand, for smaller values of $\beta$ the agreement very good.

%
\begin{figure}[t]
\includegraphics[width=0.9\linewidth]{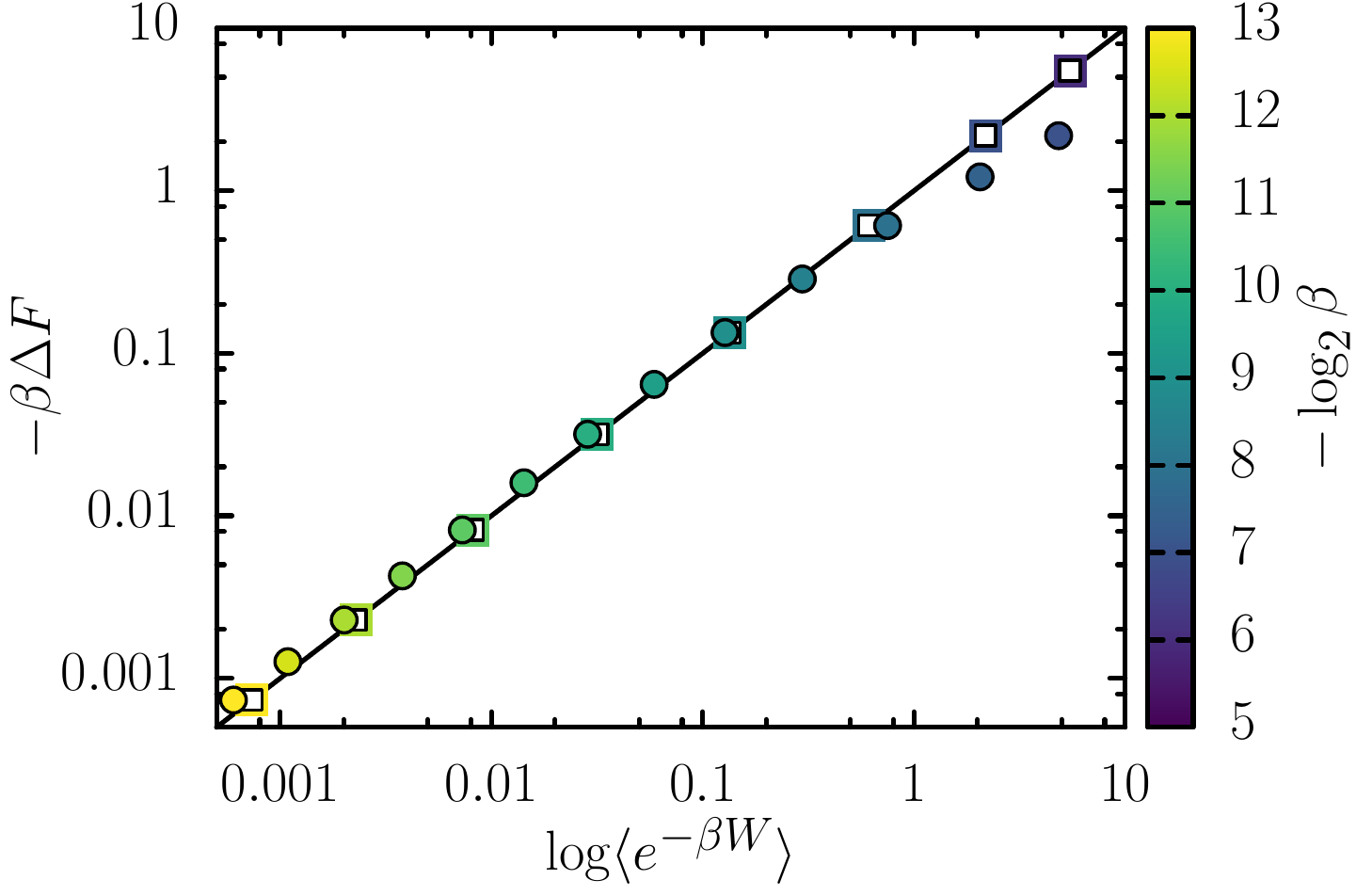} 
\caption{Jarzinsky relation for the quenched described in the text. The squares correspond to the results obtained using the quantum work distribution,
and the circles have been computed using the semiclassical approximation of the characteristic function for different temperatures. 
\label{fig:jarz} }
\end{figure}

Some of the importance of $P(W)$ comes from the quantum fluctuation relations and  the possibility, for example, 
to extract thermodynamical information like the change in free energy, from the quantum Jarzynsky relation
\begin{equation}
\label{jarz}
\Delta F\equiv-\frac{1}{\beta}\ln\frac{Z^Q_{\xi_f}}{Z^Q_{\xi_0}}=- \frac{1}{\beta}\ln\langle e^{-\beta W }\rangle
\end{equation}
in systems out of equilibrium. Here we use Jarzynksy equality to test the accuracy of the semiclassical approximation. 
In Fig.~\ref{fig:jarz} we show the evaluation of Jarzynski equaltity for a larger set of temperatures in the range $\beta^{-1}\in[2^7,2^{13}]$  
for both the quantum and semiclassical calculations. As expected, Fig.~\ref{fig:PdeW} shows that the agreement achieved by the semiclassical 
approximation is better as temperature increases. 

\begin{figure}[t]
\includegraphics[width=0.9\linewidth]{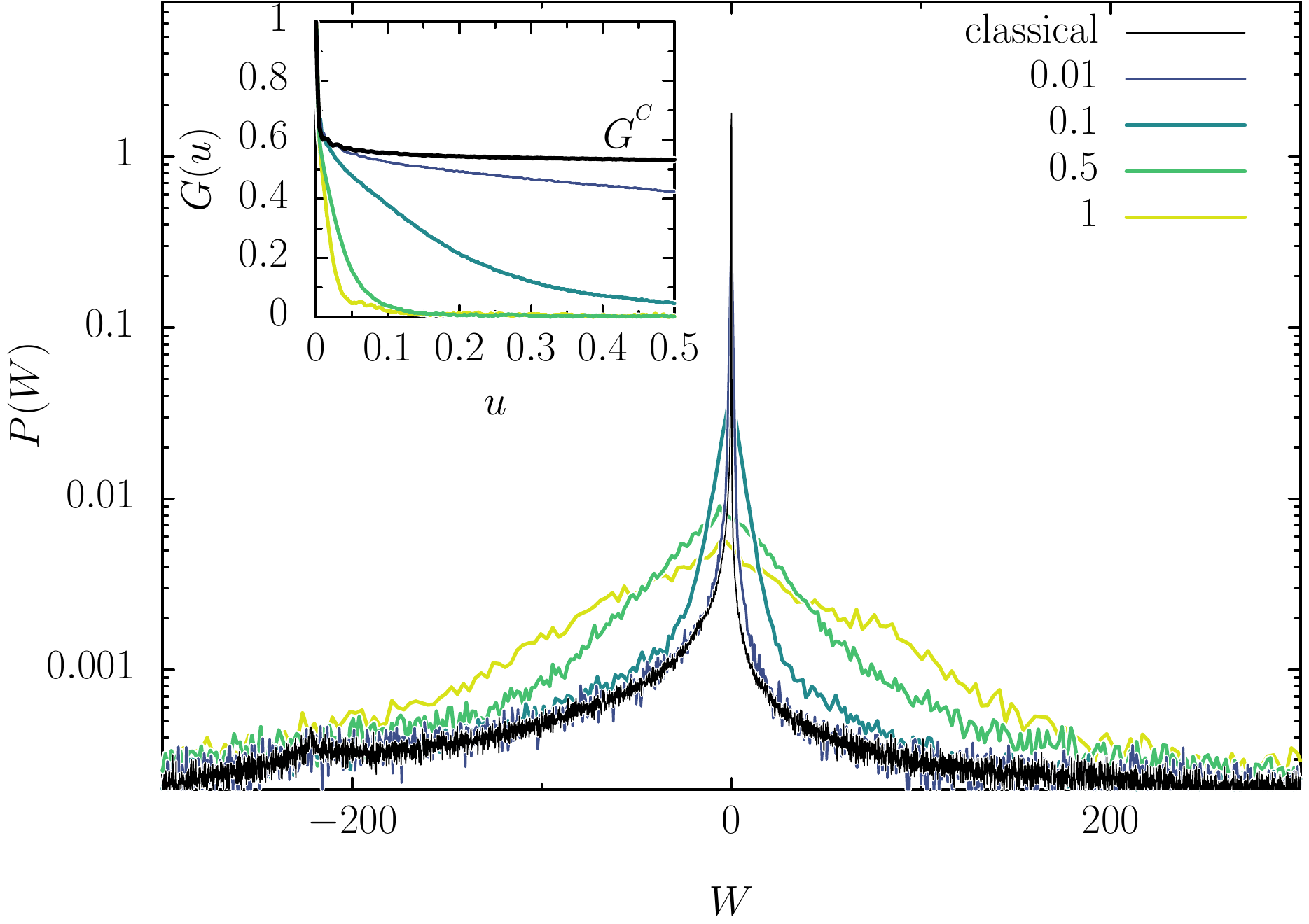}
\caption{Classical and semiclassical work distributions for $\beta~=~2^{-12}$, and 
$\hbar= 0.01, 0.1, 0.5, 1$. Semiclassical results were obtained by the numerical evaluation of Eq. \eqref{GSC}.
\label{fig:PWcl} }
\end{figure}

In Figs.~\ref{fig:PdeW} and \ref{fig:jarz} we have demonstrated that our semiclassical method provides a good approximation of $P^Q(W)$ for a chaotic system as the 
temperature increases.  Now we proceed to show that the quantum-classical correspondence is achieved by taking the limit $\hbar\to 0$ and 
comparing the resulting work distribution with its classical counterpart (see Supplementary Material \ref{appendix}).
For numerics, the classical work distribution is obtained by randomly sampling initial conditions in phase space, 
with the corresponding energies $E_{\xi_0}=p_x^2+p_y^2$. Then, from $H_{\xi_f}$ we evaluate the final energies $E_{\xi_f}$, and obtain 
the classical transition probabilities  \cite{Jarzynski2015}. Finally, $P^C(W)$ is evaluated by considering a Boltzman distribution of initial conditions.
In the main panel of Fig.~\ref{fig:PWcl} we show $P^{SC}(W)$ for $\hbar=0.01,0.1,0.5,1$ and $P^{C}(W)$ (obtained using $4\times 10^6$ initial conditions). 
In this case it is easy to see that the classical distribution of work is independent of the temperature, and as we increase the temperature and reduce the value of $\hbar$
the semiclassical distribution approaches the classical one. In the inset of Fig.~\ref{fig:PWcl} the corresponding 
characteristic function is shown for different $\hbar$ and as expected it can also be seen that $G^{SC}\stackrel{\hbar\to 0}{\longrightarrow}G^C$.

In summary, we have shown the quantum to classical correspondence of the work distribution for a sudden quench applied on a chaotic system.
This was done using a semiclassical approach that provides a bridge between quantum and classical domains. This approach is based on the study of
the characteristic function that, in the quantum case, can be expressed as a quantum evolution.
In this way, we developed a semiclassical expression for the characteristic function that relies on the dephasing representation
and on the quantum ergodic conjecture, that states that the Wigner function of chaotic eigenfunctions are equidistributed over the corresponding energy shell. 
We have shown that the work distribution obtained from the semiclassical characteristic function is in good agreement with the quantum one over
a wide range of temperatures, and its accuracy increases with the temperature. 
Finally, we showed that this semiclassical expression allows us to test the correspondence principle for high temperatures by making $\hbar\to 0$. 
To conclude, one can notice that in general the characteristic function $G^Q(u)~=~\tr[e^{i u H_{\xi_f}}U_\tau e^{-i u H_{\xi_0}} \rho_\beta U^\dagger_\tau]$, where
$U_\tau$ is the unitary operation resulting from a given process, can be also interpreted as a quench 
$G^Q(u)~=~\tr[e^{i u  (U^\dagger_\tau H_{\xi_f}U_\tau)} e^{-i u H_{\xi_0}} \rho_\beta]$, 
therefore we expect that the ideas presented in this work could also be extended to processes different from quenches \cite{GRWPrep}. \\

\noindent
\textit{Acknowledgments.}
The authors acknowledge Marcos Saraceno for useful discussions.
Finantial support from CONICET (grants PIP 114-20110100048 and 
PIP 11220080100728), ANPCyT (grants PICT-2013-0621 and PICT 2014-3711) and UBACyT.

\begin{thebibliography}{43}%
\makeatletter
\providecommand \@ifxundefined [1]{%
 \@ifx{#1\undefined}
}%
\providecommand \@ifnum [1]{%
 \ifnum #1\expandafter \@firstoftwo
 \else \expandafter \@secondoftwo
 \fi
}%
\providecommand \@ifx [1]{%
 \ifx #1\expandafter \@firstoftwo
 \else \expandafter \@secondoftwo
 \fi
}%
\providecommand \natexlab [1]{#1}%
\providecommand \enquote  [1]{``#1''}%
\providecommand \bibnamefont  [1]{#1}%
\providecommand \bibfnamefont [1]{#1}%
\providecommand \citenamefont [1]{#1}%
\providecommand \href@noop [0]{\@secondoftwo}%
\providecommand \href [0]{\begingroup \@sanitize@url \@href}%
\providecommand \@href[1]{\@@startlink{#1}\@@href}%
\providecommand \@@href[1]{\endgroup#1\@@endlink}%
\providecommand \@sanitize@url [0]{\catcode `\\12\catcode `\$12\catcode
  `\&12\catcode `\#12\catcode `\^12\catcode `\_12\catcode `\%12\relax}%
\providecommand \@@startlink[1]{}%
\providecommand \@@endlink[0]{}%
\providecommand \url  [0]{\begingroup\@sanitize@url \@url }%
\providecommand \@url [1]{\endgroup\@href {#1}{\urlprefix }}%
\providecommand \urlprefix  [0]{URL }%
\providecommand \Eprint [0]{\href }%
\providecommand \doibase [0]{http://dx.doi.org/}%
\providecommand \selectlanguage [0]{\@gobble}%
\providecommand \bibinfo  [0]{\@secondoftwo}%
\providecommand \bibfield  [0]{\@secondoftwo}%
\providecommand \translation [1]{[#1]}%
\providecommand \BibitemOpen [0]{}%
\providecommand \bibitemStop [0]{}%
\providecommand \bibitemNoStop [0]{.\EOS\space}%
\providecommand \EOS [0]{\spacefactor3000\relax}%
\providecommand \BibitemShut  [1]{\csname bibitem#1\endcsname}%
\let\auto@bib@innerbib\@empty
\bibitem [{\citenamefont {Campisi}\ \emph {et~al.}(2011)\citenamefont
  {Campisi}, \citenamefont {H\"anggi},\ and\ \citenamefont
  {Talkner}}]{Campisi2011}%
  \BibitemOpen
  \bibfield  {author} {\bibinfo {author} {\bibfnamefont {M.}~\bibnamefont
  {Campisi}}, \bibinfo {author} {\bibfnamefont {P.}~\bibnamefont {H\"anggi}}, \
  and\ \bibinfo {author} {\bibfnamefont {P.}~\bibnamefont {Talkner}},\ }\href
  {\doibase 10.1103/RevModPhys.83.771} {\bibfield  {journal} {\bibinfo
  {journal} {Rev. Mod. Phys.}\ }\textbf {\bibinfo {volume} {83}},\ \bibinfo
  {pages} {771} (\bibinfo {year} {2011})}\BibitemShut {NoStop}%
\bibitem [{\citenamefont {Goold}\ \emph {et~al.}(2016)\citenamefont {Goold},
  \citenamefont {Huber}, \citenamefont {Riera}, \citenamefont {del Rio},\ and\
  \citenamefont {Skrzypczyk}}]{Goold16}%
  \BibitemOpen
  \bibfield  {author} {\bibinfo {author} {\bibfnamefont {J.}~\bibnamefont
  {Goold}}, \bibinfo {author} {\bibfnamefont {M.}~\bibnamefont {Huber}},
  \bibinfo {author} {\bibfnamefont {A.}~\bibnamefont {Riera}}, \bibinfo
  {author} {\bibfnamefont {L.}~\bibnamefont {del Rio}}, \ and\ \bibinfo
  {author} {\bibfnamefont {P.}~\bibnamefont {Skrzypczyk}},\ }\href
  {http://stacks.iop.org/1751-8121/49/i=14/a=143001} {\bibfield  {journal}
  {\bibinfo  {journal} {J. Phys. A: Math. Theo.}\ }\textbf {\bibinfo {volume}
  {49}},\ \bibinfo {pages} {143001} (\bibinfo {year} {2016})}\BibitemShut
  {NoStop}%
\bibitem [{\citenamefont {Kurchan}(2000)}]{Kurchan2000}%
  \BibitemOpen
  \bibfield  {author} {\bibinfo {author} {\bibfnamefont {J.}~\bibnamefont
  {Kurchan}},\ }\href@noop {} {\bibfield  {journal} {\bibinfo  {journal}
  {arXiv:cond-mat/0007360}\ } (\bibinfo {year} {2000})}\BibitemShut {NoStop}%
\bibitem [{\citenamefont {Tasaki}()}]{Tasaki00}%
  \BibitemOpen
  \bibfield  {author} {\bibinfo {author} {\bibfnamefont {H.}~\bibnamefont
  {Tasaki}},\ }\href@noop {} {\bibinfo  {journal} {arXiv:cond-mat/0009244v2}\
  }\BibitemShut {NoStop}%
\bibitem [{\citenamefont {Talkner}\ \emph {et~al.}(2007)\citenamefont
  {Talkner}, \citenamefont {Lutz},\ and\ \citenamefont {H\"anggi}}]{Talkner07}%
  \BibitemOpen
\bibfield  {journal} {  }\bibfield  {author} {\bibinfo {author} {\bibfnamefont
  {P.}~\bibnamefont {Talkner}}, \bibinfo {author} {\bibfnamefont
  {E.}~\bibnamefont {Lutz}}, \ and\ \bibinfo {author} {\bibfnamefont
  {P.}~\bibnamefont {H\"anggi}},\ }\href {\doibase 10.1103/PhysRevE.75.050102}
  {\bibfield  {journal} {\bibinfo  {journal} {Phys. Rev. E}\ }\textbf {\bibinfo
  {volume} {75}},\ \bibinfo {pages} {050102} (\bibinfo {year}
  {2007})}\BibitemShut {NoStop}%
\bibitem [{\citenamefont {Jarzynski}(1997)}]{Jarzynski1997}%
  \BibitemOpen
  \bibfield  {author} {\bibinfo {author} {\bibfnamefont {C.}~\bibnamefont
  {Jarzynski}},\ }\href {\doibase 10.1103/PhysRevLett.78.2690} {\bibfield
  {journal} {\bibinfo  {journal} {Phys. Rev. Lett.}\ }\textbf {\bibinfo
  {volume} {78}},\ \bibinfo {pages} {2690} (\bibinfo {year}
  {1997})}\BibitemShut {NoStop}%
\bibitem [{\citenamefont {Crooks}(1999)}]{Crooks1999}%
  \BibitemOpen
  \bibfield  {author} {\bibinfo {author} {\bibfnamefont {G.~E.}\ \bibnamefont
  {Crooks}},\ }\href {\doibase 10.1103/PhysRevE.60.2721} {\bibfield  {journal}
  {\bibinfo  {journal} {Phys. Rev. E}\ }\textbf {\bibinfo {volume} {60}},\
  \bibinfo {pages} {2721} (\bibinfo {year} {1999})}\BibitemShut {NoStop}%
\bibitem [{\citenamefont {Jarzynski}\ \emph {et~al.}(2015)\citenamefont
  {Jarzynski}, \citenamefont {Quan},\ and\ \citenamefont
  {Rahav}}]{Jarzynski2015}%
  \BibitemOpen
  \bibfield  {author} {\bibinfo {author} {\bibfnamefont {C.}~\bibnamefont
  {Jarzynski}}, \bibinfo {author} {\bibfnamefont {H.~T.}\ \bibnamefont {Quan}},
  \ and\ \bibinfo {author} {\bibfnamefont {S.}~\bibnamefont {Rahav}},\ }\href
  {\doibase 10.1103/PhysRevX.5.031038} {\bibfield  {journal} {\bibinfo
  {journal} {Phys. Rev. X}\ }\textbf {\bibinfo {volume} {5}},\ \bibinfo {pages}
  {031038} (\bibinfo {year} {2015})}\BibitemShut {NoStop}%
\bibitem [{\citenamefont {Zhu}\ \emph {et~al.}(2016)\citenamefont {Zhu},
  \citenamefont {Gong}, \citenamefont {Wu},\ and\ \citenamefont
  {Quan}}]{Quan2016}%
  \BibitemOpen
  \bibfield  {author} {\bibinfo {author} {\bibfnamefont {L.}~\bibnamefont
  {Zhu}}, \bibinfo {author} {\bibfnamefont {Z.}~\bibnamefont {Gong}}, \bibinfo
  {author} {\bibfnamefont {B.}~\bibnamefont {Wu}}, \ and\ \bibinfo {author}
  {\bibfnamefont {H.~T.}\ \bibnamefont {Quan}},\ }\href {\doibase
  10.1103/PhysRevE.93.062108} {\bibfield  {journal} {\bibinfo  {journal} {Phys.
  Rev. E}\ }\textbf {\bibinfo {volume} {93}},\ \bibinfo {pages} {062108}
  (\bibinfo {year} {2016})}\BibitemShut {NoStop}%
\bibitem [{\citenamefont {Deffner}\ and\ \citenamefont
  {Lutz}(2008{\natexlab{a}})}]{Deffner08}%
  \BibitemOpen
  \bibfield  {author} {\bibinfo {author} {\bibfnamefont {S.}~\bibnamefont
  {Deffner}}\ and\ \bibinfo {author} {\bibfnamefont {E.}~\bibnamefont {Lutz}},\
  }\href {\doibase 10.1103/PhysRevE.77.021128} {\bibfield  {journal} {\bibinfo
  {journal} {Phys. Rev. E}\ }\textbf {\bibinfo {volume} {77}},\ \bibinfo
  {pages} {021128} (\bibinfo {year} {2008}{\natexlab{a}})}\BibitemShut
  {NoStop}%
\bibitem [{\citenamefont {Talkner}\ \emph {et~al.}(2008)\citenamefont
  {Talkner}, \citenamefont {Burada},\ and\ \citenamefont
  {H\"anggi}}]{Talkner08}%
  \BibitemOpen
  \bibfield  {author} {\bibinfo {author} {\bibfnamefont {P.}~\bibnamefont
  {Talkner}}, \bibinfo {author} {\bibfnamefont {P.~S.}\ \bibnamefont {Burada}},
  \ and\ \bibinfo {author} {\bibfnamefont {P.}~\bibnamefont {H\"anggi}},\
  }\href {\doibase 10.1103/PhysRevE.78.011115} {\bibfield  {journal} {\bibinfo
  {journal} {Phys. Rev. E}\ }\textbf {\bibinfo {volume} {78}},\ \bibinfo
  {pages} {011115} (\bibinfo {year} {2008})}\BibitemShut {NoStop}%
\bibitem [{\citenamefont {Campisi}(2008)}]{Campisi08}%
  \BibitemOpen
  \bibfield  {author} {\bibinfo {author} {\bibfnamefont {M.}~\bibnamefont
  {Campisi}},\ }\href {\doibase 10.1103/PhysRevE.78.051123} {\bibfield
  {journal} {\bibinfo  {journal} {Phys. Rev. E}\ }\textbf {\bibinfo {volume}
  {78}},\ \bibinfo {pages} {051123} (\bibinfo {year} {2008})}\BibitemShut
  {NoStop}%
\bibitem [{\citenamefont {Deffner}\ \emph {et~al.}(2010)\citenamefont
  {Deffner}, \citenamefont {Abah},\ and\ \citenamefont {Lutz}}]{Deffner10}%
  \BibitemOpen
  \bibfield  {author} {\bibinfo {author} {\bibfnamefont {S.}~\bibnamefont
  {Deffner}}, \bibinfo {author} {\bibfnamefont {O.}~\bibnamefont {Abah}}, \
  and\ \bibinfo {author} {\bibfnamefont {E.}~\bibnamefont {Lutz}},\ }\href
  {\doibase http://dx.doi.org/10.1016/j.chemphys.2010.04.042} {\bibfield
  {journal} {\bibinfo  {journal} {Chemical Physics}\ }\textbf {\bibinfo
  {volume} {375}},\ \bibinfo {pages} {200 } (\bibinfo {year}
  {2010})}\BibitemShut {NoStop}%
\bibitem [{\citenamefont {Ford}\ \emph {et~al.}(2012)\citenamefont {Ford},
  \citenamefont {Minor},\ and\ \citenamefont {Binnie}}]{Ford12}%
  \BibitemOpen
  \bibfield  {author} {\bibinfo {author} {\bibfnamefont {I.~J.}\ \bibnamefont
  {Ford}}, \bibinfo {author} {\bibfnamefont {D.~S.}\ \bibnamefont {Minor}}, \
  and\ \bibinfo {author} {\bibfnamefont {S.~J.}\ \bibnamefont {Binnie}},\
  }\href {http://stacks.iop.org/0143-0807/33/i=6/a=1789} {\bibfield  {journal}
  {\bibinfo  {journal} {European Journal of Physics}\ }\textbf {\bibinfo
  {volume} {33}},\ \bibinfo {pages} {1789} (\bibinfo {year}
  {2012})}\BibitemShut {NoStop}%
\bibitem [{\citenamefont {Talkner}\ \emph {et~al.}(2013)\citenamefont
  {Talkner}, \citenamefont {Morillo}, \citenamefont {Yi},\ and\ \citenamefont
  {Hänggi}}]{Talkner13}%
  \BibitemOpen
  \bibfield  {author} {\bibinfo {author} {\bibfnamefont {P.}~\bibnamefont
  {Talkner}}, \bibinfo {author} {\bibfnamefont {M.}~\bibnamefont {Morillo}},
  \bibinfo {author} {\bibfnamefont {J.}~\bibnamefont {Yi}}, \ and\ \bibinfo
  {author} {\bibfnamefont {P.}~\bibnamefont {Hänggi}},\ }\href
  {http://stacks.iop.org/1367-2630/15/i=9/a=095001} {\bibfield  {journal}
  {\bibinfo  {journal} {New Journal of Physics}\ }\textbf {\bibinfo {volume}
  {15}},\ \bibinfo {pages} {095001} (\bibinfo {year} {2013})}\BibitemShut
  {NoStop}%
\bibitem [{\citenamefont {Silva}(2008)}]{Silva2008}%
  \BibitemOpen
  \bibfield  {author} {\bibinfo {author} {\bibfnamefont {A.}~\bibnamefont
  {Silva}},\ }\href@noop {} {\bibfield  {journal} {\bibinfo  {journal} {Phys.
  Rev. Lett.}\ }\textbf {\bibinfo {volume} {101}},\ \bibinfo {pages} {120603}
  (\bibinfo {year} {2008})}\BibitemShut {NoStop}%
\bibitem [{\citenamefont {Van\'i\ifmmode~\check{c}\else \v{c}\fi{}ek}\ and\
  \citenamefont {Heller}(2003)}]{vanicek2003}%
  \BibitemOpen
  \bibfield  {author} {\bibinfo {author} {\bibfnamefont {J.}~\bibnamefont
  {Van\'i\ifmmode~\check{c}\else \v{c}\fi{}ek}}\ and\ \bibinfo {author}
  {\bibfnamefont {E.~J.}\ \bibnamefont {Heller}},\ }\href@noop {} {\bibfield
  {journal} {\bibinfo  {journal} {Phys. Rev. E}\ }\textbf {\bibinfo {volume}
  {68}},\ \bibinfo {pages} {056208} (\bibinfo {year} {2003})}\BibitemShut
  {NoStop}%
\bibitem [{\citenamefont {Van\'i\ifmmode~\check{c}\else
  \v{c}\fi{}ek}(2004)}]{vanicek2004}%
  \BibitemOpen
  \bibfield  {author} {\bibinfo {author} {\bibfnamefont {J.}~\bibnamefont
  {Van\'i\ifmmode~\check{c}\else \v{c}\fi{}ek}},\ }\href@noop {} {\bibfield
  {journal} {\bibinfo  {journal} {Phys. Rev. E}\ }\textbf {\bibinfo {volume}
  {70}},\ \bibinfo {pages} {055201 (R)} (\bibinfo {year} {2004})}\BibitemShut
  {NoStop}%
\bibitem [{\citenamefont {Van{\'\i}{\v c}ek}(2006)}]{vanicek2006}%
  \BibitemOpen
  \bibfield  {author} {\bibinfo {author} {\bibfnamefont {J.}~\bibnamefont
  {Van{\'\i}{\v c}ek}},\ }\href@noop {} {\bibfield  {journal} {\bibinfo
  {journal} {Phys. Rev. E}\ }\textbf {\bibinfo {volume} {73}},\ \bibinfo
  {pages} {046204} (\bibinfo {year} {2006})}\BibitemShut {NoStop}%
\bibitem [{\citenamefont {Berry}(1977)}]{Berry_1977}%
  \BibitemOpen
  \bibfield  {author} {\bibinfo {author} {\bibfnamefont {M.}~\bibnamefont
  {Berry}},\ }\href@noop {} {\bibfield  {journal} {\bibinfo  {journal} {J.
  Phys.~A: Math. Gen.}\ }\textbf {\bibinfo {volume} {10}},\ \bibinfo {pages}
  {2083} (\bibinfo {year} {1977})}\BibitemShut {NoStop}%
\bibitem [{\citenamefont {Voros}(1979)}]{Voros}%
  \BibitemOpen
  \bibfield  {author} {\bibinfo {author} {\bibfnamefont {A.}~\bibnamefont
  {Voros}},\ }\href@noop {} {\emph {\bibinfo {title} {in Stochastic behavior in
  classical and quantum Hamiltonian systems}}},\ edited by\ \bibinfo {editor}
  {\bibfnamefont {M.-J.}\ \bibnamefont {Giannoni}}, \bibinfo {editor}
  {\bibfnamefont {A.}~\bibnamefont {Voros}}, \ and\ \bibinfo {editor}
  {\bibfnamefont {J.}~\bibnamefont {Zinn-Justin}}\ (\bibinfo  {publisher}
  {{Lectures notes in Physics 93. Berlin: Springer }},\ \bibinfo {year}
  {1979})\BibitemShut {NoStop}%
\bibitem [{\citenamefont {Huber}\ \emph {et~al.}(2008)\citenamefont {Huber},
  \citenamefont {Schmidt-Kaler}, \citenamefont {Deffner},\ and\ \citenamefont
  {Lutz}}]{Huber2008}%
  \BibitemOpen
  \bibfield  {author} {\bibinfo {author} {\bibfnamefont {G.}~\bibnamefont
  {Huber}}, \bibinfo {author} {\bibfnamefont {F.}~\bibnamefont
  {Schmidt-Kaler}}, \bibinfo {author} {\bibfnamefont {S.}~\bibnamefont
  {Deffner}}, \ and\ \bibinfo {author} {\bibfnamefont {E.}~\bibnamefont
  {Lutz}},\ }\href {\doibase 10.1103/PhysRevLett.101.070403} {\bibfield
  {journal} {\bibinfo  {journal} {Phys. Rev. Lett.}\ }\textbf {\bibinfo
  {volume} {101}},\ \bibinfo {pages} {070403} (\bibinfo {year}
  {2008})}\BibitemShut {NoStop}%
\bibitem [{\citenamefont {Dorner}\ \emph {et~al.}(2013)\citenamefont {Dorner},
  \citenamefont {Clark}, \citenamefont {Heaney}, \citenamefont {Fazio},
  \citenamefont {Goold},\ and\ \citenamefont {Vedral}}]{Dorner2013}%
  \BibitemOpen
  \bibfield  {author} {\bibinfo {author} {\bibfnamefont {R.}~\bibnamefont
  {Dorner}}, \bibinfo {author} {\bibfnamefont {S.~R.}\ \bibnamefont {Clark}},
  \bibinfo {author} {\bibfnamefont {L.}~\bibnamefont {Heaney}}, \bibinfo
  {author} {\bibfnamefont {R.}~\bibnamefont {Fazio}}, \bibinfo {author}
  {\bibfnamefont {J.}~\bibnamefont {Goold}}, \ and\ \bibinfo {author}
  {\bibfnamefont {V.}~\bibnamefont {Vedral}},\ }\href {\doibase
  10.1103/PhysRevLett.110.230601} {\bibfield  {journal} {\bibinfo  {journal}
  {Phys. Rev. Lett.}\ }\textbf {\bibinfo {volume} {110}},\ \bibinfo {pages}
  {230601} (\bibinfo {year} {2013})}\BibitemShut {NoStop}%
\bibitem [{\citenamefont {Mazzola}\ \emph {et~al.}(2013)\citenamefont
  {Mazzola}, \citenamefont {De~Chiara},\ and\ \citenamefont
  {Paternostro}}]{Mazzola2013}%
  \BibitemOpen
  \bibfield  {author} {\bibinfo {author} {\bibfnamefont {L.}~\bibnamefont
  {Mazzola}}, \bibinfo {author} {\bibfnamefont {G.}~\bibnamefont {De~Chiara}},
  \ and\ \bibinfo {author} {\bibfnamefont {M.}~\bibnamefont {Paternostro}},\
  }\href@noop {} {\bibfield  {journal} {\bibinfo  {journal} {Phys. Rev. Lett.}\
  }\textbf {\bibinfo {volume} {110}},\ \bibinfo {pages} {230602} (\bibinfo
  {year} {2013})}\BibitemShut {NoStop}%
\bibitem [{\citenamefont {Roncaglia}\ \emph {et~al.}(2014)\citenamefont
  {Roncaglia}, \citenamefont {Cerisola},\ and\ \citenamefont
  {Paz}}]{Roncaglia2014}%
  \BibitemOpen
  \bibfield  {author} {\bibinfo {author} {\bibfnamefont {A.~J.}\ \bibnamefont
  {Roncaglia}}, \bibinfo {author} {\bibfnamefont {F.}~\bibnamefont {Cerisola}},
  \ and\ \bibinfo {author} {\bibfnamefont {J.~P.}\ \bibnamefont {Paz}},\ }\href
  {\doibase 10.1103/PhysRevLett.113.250601} {\bibfield  {journal} {\bibinfo
  {journal} {Phys. Rev. Lett.}\ }\textbf {\bibinfo {volume} {113}},\ \bibinfo
  {pages} {250601} (\bibinfo {year} {2014})}\BibitemShut {NoStop}%
\bibitem [{\citenamefont {Batalh{\~a}o}\ \emph {et~al.}(2014)\citenamefont
  {Batalh{\~a}o}, \citenamefont {Souza}, \citenamefont {Mazzola}, \citenamefont
  {Auccaise}, \citenamefont {Sarthour}, \citenamefont {Oliveira}, \citenamefont
  {Goold}, \citenamefont {De~Chiara}, \citenamefont {Paternostro},\ and\
  \citenamefont {Serra}}]{Batalhao2014}%
  \BibitemOpen
  \bibfield  {author} {\bibinfo {author} {\bibfnamefont {T.~B.}\ \bibnamefont
  {Batalh{\~a}o}}, \bibinfo {author} {\bibfnamefont {A.~M.}\ \bibnamefont
  {Souza}}, \bibinfo {author} {\bibfnamefont {L.}~\bibnamefont {Mazzola}},
  \bibinfo {author} {\bibfnamefont {R.}~\bibnamefont {Auccaise}}, \bibinfo
  {author} {\bibfnamefont {R.~S.}\ \bibnamefont {Sarthour}}, \bibinfo {author}
  {\bibfnamefont {I.~S.}\ \bibnamefont {Oliveira}}, \bibinfo {author}
  {\bibfnamefont {J.}~\bibnamefont {Goold}}, \bibinfo {author} {\bibfnamefont
  {G.}~\bibnamefont {De~Chiara}}, \bibinfo {author} {\bibfnamefont
  {M.}~\bibnamefont {Paternostro}}, \ and\ \bibinfo {author} {\bibfnamefont
  {R.~M.}\ \bibnamefont {Serra}},\ }\href@noop {} {\bibfield  {journal}
  {\bibinfo  {journal} {Phys. Rev. Lett.}\ }\textbf {\bibinfo {volume} {113}},\
  \bibinfo {pages} {140601} (\bibinfo {year} {2014})}\BibitemShut {NoStop}%
\bibitem [{\citenamefont {An}\ \emph {et~al.}(2014)\citenamefont {An},
  \citenamefont {Zhang}, \citenamefont {Um}, \citenamefont {Lv}, \citenamefont
  {Lu}, \citenamefont {Zhang}, \citenamefont {Yin}, \citenamefont {Quan},\ and\
  \citenamefont {Kim}}]{An2014}%
  \BibitemOpen
  \bibfield  {author} {\bibinfo {author} {\bibfnamefont {S.}~\bibnamefont
  {An}}, \bibinfo {author} {\bibfnamefont {J.-N.}\ \bibnamefont {Zhang}},
  \bibinfo {author} {\bibfnamefont {M.}~\bibnamefont {Um}}, \bibinfo {author}
  {\bibfnamefont {D.}~\bibnamefont {Lv}}, \bibinfo {author} {\bibfnamefont
  {Y.}~\bibnamefont {Lu}}, \bibinfo {author} {\bibfnamefont {J.}~\bibnamefont
  {Zhang}}, \bibinfo {author} {\bibfnamefont {Z.-Q.}\ \bibnamefont {Yin}},
  \bibinfo {author} {\bibfnamefont {H.~T.}\ \bibnamefont {Quan}}, \ and\
  \bibinfo {author} {\bibfnamefont {K.}~\bibnamefont {Kim}},\ }\href@noop {}
  {\bibfield  {journal} {\bibinfo  {journal} {Nature Physics}\ }\textbf
  {\bibinfo {volume} {11}},\ \bibinfo {pages} {193} (\bibinfo {year}
  {2014})}\BibitemShut {NoStop}%
\bibitem [{\citenamefont {Deffner}\ and\ \citenamefont
  {Lutz}(2008{\natexlab{b}})}]{Deffner2008}%
  \BibitemOpen
  \bibfield  {author} {\bibinfo {author} {\bibfnamefont {S.}~\bibnamefont
  {Deffner}}\ and\ \bibinfo {author} {\bibfnamefont {E.}~\bibnamefont {Lutz}},\
  }\href {\doibase 10.1103/PhysRevE.77.021128} {\bibfield  {journal} {\bibinfo
  {journal} {Phys. Rev. E}\ }\textbf {\bibinfo {volume} {77}},\ \bibinfo
  {pages} {021128} (\bibinfo {year} {2008}{\natexlab{b}})}\BibitemShut
  {NoStop}%
\bibitem [{\citenamefont {Gorin}\ \emph {et~al.}(2006)\citenamefont {Gorin},
  \citenamefont {Prosen}, \citenamefont {Seligman},\ and\ \citenamefont {{\v
  Z}nidari{\v c}}}]{Gorin2006}%
  \BibitemOpen
  \bibfield  {author} {\bibinfo {author} {\bibfnamefont {T.}~\bibnamefont
  {Gorin}}, \bibinfo {author} {\bibfnamefont {T.}~\bibnamefont {Prosen}},
  \bibinfo {author} {\bibfnamefont {T.}~\bibnamefont {Seligman}}, \ and\
  \bibinfo {author} {\bibfnamefont {M.}~\bibnamefont {{\v Z}nidari{\v c}}},\
  }\href@noop {} {\bibfield  {journal} {\bibinfo  {journal} {Phys. Rep.}\
  }\textbf {\bibinfo {volume} {435}},\ \bibinfo {pages} {33} (\bibinfo {year}
  {2006})}\BibitemShut {NoStop}%
\bibitem [{\citenamefont {\mbox{Ph.} Jacquod}\ and\ \citenamefont
  {Petitjean}(2009)}]{Jacquod2009}%
  \BibitemOpen
  \bibfield  {author} {\bibinfo {author} {\bibnamefont {\mbox{Ph.} Jacquod}}\
  and\ \bibinfo {author} {\bibfnamefont {C.}~\bibnamefont {Petitjean}},\
  }\href@noop {} {\bibfield  {journal} {\bibinfo  {journal} {Adv. Phys.}\
  }\textbf {\bibinfo {volume} {58}},\ \bibinfo {pages} {67} (\bibinfo {year}
  {2009})}\BibitemShut {NoStop}%
\bibitem [{\citenamefont {Goussev}\ \emph {et~al.}(2012)\citenamefont
  {Goussev}, \citenamefont {Jalabert}, \citenamefont {Pastawski},\ and\
  \citenamefont {Wisniacki}}]{DiegoScholar}%
  \BibitemOpen
  \bibfield  {author} {\bibinfo {author} {\bibfnamefont {A.}~\bibnamefont
  {Goussev}}, \bibinfo {author} {\bibfnamefont {R.}~\bibnamefont {Jalabert}},
  \bibinfo {author} {\bibfnamefont {H.~M.}\ \bibnamefont {Pastawski}}, \ and\
  \bibinfo {author} {\bibfnamefont {D.~A.}\ \bibnamefont {Wisniacki}},\
  }\href@noop {} {\bibfield  {journal} {\bibinfo  {journal} {Scholarpedia}\
  }\textbf {\bibinfo {volume} {7}},\ \bibinfo {pages} {11687} (\bibinfo {year}
  {2012})}\BibitemShut {NoStop}%
\bibitem [{\citenamefont {Miller}(1970)}]{Miller1970}%
  \BibitemOpen
  \bibfield  {author} {\bibinfo {author} {\bibfnamefont {W.~H.}\ \bibnamefont
  {Miller}},\ }\href@noop {} {\bibfield  {journal} {\bibinfo  {journal} {J.
  Chem. Phys.}\ }\textbf {\bibinfo {volume} {53}},\ \bibinfo {pages} {3578}
  (\bibinfo {year} {1970})}\BibitemShut {NoStop}%
\bibitem [{\citenamefont {Miller}(2001)}]{Miller2001}%
  \BibitemOpen
  \bibfield  {author} {\bibinfo {author} {\bibfnamefont {W.~H.}\ \bibnamefont
  {Miller}},\ }\href@noop {} {\bibfield  {journal} {\bibinfo  {journal} {J.
  Phys. Chem}\ }\textbf {\bibinfo {volume} {105}},\ \bibinfo {pages} {2942}
  (\bibinfo {year} {2001})}\BibitemShut {NoStop}%
\bibitem [{\citenamefont {Garc\'ia-Mata}\ \emph {et~al.}(2011)\citenamefont
  {Garc\'ia-Mata}, \citenamefont {Vallejos},\ and\ \citenamefont
  {Wisniacki}}]{garciamaNJP}%
  \BibitemOpen
  \bibfield  {author} {\bibinfo {author} {\bibfnamefont {I.}~\bibnamefont
  {Garc\'ia-Mata}}, \bibinfo {author} {\bibfnamefont {R.~O.}\ \bibnamefont
  {Vallejos}}, \ and\ \bibinfo {author} {\bibfnamefont {D.~A.}\ \bibnamefont
  {Wisniacki}},\ }\href@noop {} {\bibfield  {journal} {\bibinfo  {journal} {New
  J. Phys.}\ }\textbf {\bibinfo {volume} {13}},\ \bibinfo {pages} {103040}
  (\bibinfo {year} {2011})}\BibitemShut {NoStop}%
\bibitem [{\citenamefont {Garc{\'\i}a-Mata}\ \emph {et~al.}(2016)\citenamefont
  {Garc{\'\i}a-Mata}, \citenamefont {Roncaglia},\ and\ \citenamefont
  {Wisniacki}}]{GRW2016}%
  \BibitemOpen
  \bibfield  {author} {\bibinfo {author} {\bibfnamefont {I.}~\bibnamefont
  {Garc{\'\i}a-Mata}}, \bibinfo {author} {\bibfnamefont {A.~J.}\ \bibnamefont
  {Roncaglia}}, \ and\ \bibinfo {author} {\bibfnamefont {D.~A.}\ \bibnamefont
  {Wisniacki}},\ }\href@noop {} {\bibfield  {journal} {\bibinfo  {journal}
  {Phil.Trans. R. Soc. A}\ }\textbf {\bibinfo {volume} {374}} (\bibinfo {year}
  {2016})}\BibitemShut {NoStop}%
\bibitem [{\citenamefont {Wisniacki}\ \emph {et~al.}(2010)\citenamefont
  {Wisniacki}, \citenamefont {Ares},\ and\ \citenamefont
  {Vergini}}]{Diego2010}%
  \BibitemOpen
  \bibfield  {author} {\bibinfo {author} {\bibfnamefont {D.~A.}\ \bibnamefont
  {Wisniacki}}, \bibinfo {author} {\bibfnamefont {N.}~\bibnamefont {Ares}}, \
  and\ \bibinfo {author} {\bibfnamefont {E.~G.}\ \bibnamefont {Vergini}},\
  }\href@noop {} {\bibfield  {journal} {\bibinfo  {journal} {Phys. Rev. Lett.}\
  }\textbf {\bibinfo {volume} {104}},\ \bibinfo {pages} {254101} (\bibinfo
  {year} {2010})}\BibitemShut {NoStop}%
\bibitem [{\citenamefont {Zimmermann}\ and\ \citenamefont {Van{\'\i}{\v
  c}ek}(2010)}]{Zimm2010}%
  \BibitemOpen
  \bibfield  {author} {\bibinfo {author} {\bibfnamefont {T.}~\bibnamefont
  {Zimmermann}}\ and\ \bibinfo {author} {\bibfnamefont {J.}~\bibnamefont
  {Van{\'\i}{\v c}ek}},\ }\href@noop {} {\bibfield  {journal} {\bibinfo
  {journal} {J. Chem. Phys.}\ }\textbf {\bibinfo {volume} {132}},\ \bibinfo
  {pages} {241101} (\bibinfo {year} {2010})}\BibitemShut {NoStop}%
\bibitem [{\citenamefont {Zimmermann}\ \emph {et~al.}(2010)\citenamefont
  {Zimmermann}, \citenamefont {Ruppen}, \citenamefont {Li},\ and\ \citenamefont
  {Van{\'\i}{\v c}ek}}]{Zimm2010_2}%
  \BibitemOpen
  \bibfield  {author} {\bibinfo {author} {\bibfnamefont {T.}~\bibnamefont
  {Zimmermann}}, \bibinfo {author} {\bibfnamefont {J.}~\bibnamefont {Ruppen}},
  \bibinfo {author} {\bibfnamefont {B.}~\bibnamefont {Li}}, \ and\ \bibinfo
  {author} {\bibfnamefont {J.}~\bibnamefont {Van{\'\i}{\v c}ek}},\ }\href@noop
  {} {\bibfield  {journal} {\bibinfo  {journal} {Int. J. Quant. Chem.}\
  }\textbf {\bibinfo {volume} {110}},\ \bibinfo {pages} {2426} (\bibinfo {year}
  {2010})}\BibitemShut {NoStop}%
\bibitem [{\citenamefont {Schwieters}\ and\ \citenamefont
  {Delos}(1995)}]{Schwieters95}%
  \BibitemOpen
  \bibfield  {author} {\bibinfo {author} {\bibfnamefont {C.~D.}\ \bibnamefont
  {Schwieters}}\ and\ \bibinfo {author} {\bibfnamefont {J.~B.}\ \bibnamefont
  {Delos}},\ }\href {\doibase 10.1103/PhysRevA.51.1030} {\bibfield  {journal}
  {\bibinfo  {journal} {Phys. Rev. A}\ }\textbf {\bibinfo {volume} {51}},\
  \bibinfo {pages} {1030} (\bibinfo {year} {1995})}\BibitemShut {NoStop}%
\bibitem [{\citenamefont {Quan}\ and\ \citenamefont
  {Jarzynski}(2012)}]{Quan2012}%
  \BibitemOpen
  \bibfield  {author} {\bibinfo {author} {\bibfnamefont {H.~T.}\ \bibnamefont
  {Quan}}\ and\ \bibinfo {author} {\bibfnamefont {C.}~\bibnamefont
  {Jarzynski}},\ }\href {\doibase 10.1103/PhysRevE.85.031102} {\bibfield
  {journal} {\bibinfo  {journal} {Phys. Rev. E}\ }\textbf {\bibinfo {volume}
  {85}},\ \bibinfo {pages} {031102} (\bibinfo {year} {2012})}\BibitemShut
  {NoStop}%
\bibitem [{\citenamefont {Vergini}\ and\ \citenamefont
  {Saraceno}(1995)}]{vergini95}%
  \BibitemOpen
  \bibfield  {author} {\bibinfo {author} {\bibfnamefont {E.}~\bibnamefont
  {Vergini}}\ and\ \bibinfo {author} {\bibfnamefont {M.}~\bibnamefont
  {Saraceno}},\ }\href {\doibase 10.1103/PhysRevE.52.2204} {\bibfield
  {journal} {\bibinfo  {journal} {Phys. Rev. E}\ }\textbf {\bibinfo {volume}
  {52}},\ \bibinfo {pages} {2204} (\bibinfo {year} {1995})}\BibitemShut
  {NoStop}%
\bibitem [{\citenamefont {Zambrano}\ \emph {et~al.}(2013)\citenamefont
  {Zambrano}, \citenamefont {{\v S}ulc},\ and\ \citenamefont {Van{\'\i}{\v
  c}ek}}]{Zambrano2013}%
  \BibitemOpen
  \bibfield  {author} {\bibinfo {author} {\bibfnamefont {E.}~\bibnamefont
  {Zambrano}}, \bibinfo {author} {\bibfnamefont {M.}~\bibnamefont {{\v S}ulc}},
  \ and\ \bibinfo {author} {\bibfnamefont {J.}~\bibnamefont {Van{\'\i}{\v
  c}ek}},\ }\href@noop {} {\bibfield  {journal} {\bibinfo  {journal} {The
  Journal of Chemical Physics}\ }\textbf {\bibinfo {volume} {139}},\ \bibinfo
  {eid} {054109} (\bibinfo {year} {2013})}\BibitemShut {NoStop}%
\bibitem [{\citenamefont {Garc\'ia-Mata}\ \emph {et~al.}()\citenamefont
  {Garc\'ia-Mata}, \citenamefont {Roncaglia},\ and\ \citenamefont
  {Wisniacki}}]{GRWPrep}%
  \BibitemOpen
  \bibfield  {author} {\bibinfo {author} {\bibfnamefont {I.}~\bibnamefont
  {Garc\'ia-Mata}}, \bibinfo {author} {\bibfnamefont {A.~J.}\ \bibnamefont
  {Roncaglia}}, \ and\ \bibinfo {author} {\bibfnamefont {D.~A.}\ \bibnamefont
  {Wisniacki}},\ }\href@noop {} {\bibinfo  {journal} {In preparation}\
  }\BibitemShut {NoStop}%
\end{thebibliography}
%
\newpage

\onecolumngrid
\appendix
\section{SUPPLEMENTARY MATERIAL}
\label{appendix}
\subsection{Classical work distribution}
Here we establish the notation and review the definition of the classical work distribution for a quench where $H_{\xi_0}\rightarrow H_{\xi_f}$.
The classical work distribution is defined by \cite{Schwieters95,Jarzynski2015}:
\beq
P^C(W)=\int dE_{\xi_f}\int dE_{\xi_0}\bar P^C (E_{\xi_f}|E_{\xi_0}) \bar P_{\xi_0}^C(E_{\xi_0}) \delta(W-E_{\xi_f}-E_{\xi_0})
\eeq
where $\bar P^C_{\xi_0}(E_{\xi_0})$ is the classical distribution of initial energies $E_{\xi_0}$, sampled from equilibrium:
\beq
\bar P^C_{\xi_0}(E)=\frac{1}{Z_{\xi_0}^C}e^{-\beta E}g_{\xi_0}(E),
\eeq
where 
\beq
Z_\xi^C=\int dx \exp[-\beta H_\xi(x)], \;\; g_\xi(E)=\int dx \;\delta[E- H_\xi(x)],
\eeq
are the classical partition function and the density of states. Note that in these definitions we do not consider the factor $h$ for simplicity, 
as it was done in \cite{Jarzynski2015}. The conditional energy distribution for a quench is defined as:
\beq
\bar P^C (E_{\xi_f}|E_{\xi_0}) =\frac{\int dx_0\,\delta[E_{\xi_0}-H_{\xi_0}(x_0)] \; \delta [E_{\xi_f} -  H_{\xi_f}(x_0)]}{\int dx_0\; \delta[E_{\xi_0}-H_{\xi_0}(x_0)]}.
\eeq

\subsection{Semiclassical approach to the Characteristic function}

Here we sketch the derivation of a semiclassical approximation of the characteristic function. First we will show some definitions of the dephasing 
representation, and then we will show the derivation of the semiclassical expression for the characteristic function.

The Dephasing Representation (DR) is a semiclassical approximation to a general fidelity amplitude:
\beq
f(t)=\tr[ e^{i  H_{\xi_f} t/\hbar} e^{-i H_{\xi_0} t/\hbar} \rho].
\eeq
Thus, in the DR the fidelity amplitude is:
\beq
f_{DR}(t)=\int {d^{2D}x_0} \, {\cal W_\rho}(x_0) \, \exp\left[{\frac{i}{\hbar}\int_0^t \Delta H\left(x(s)\right) ds}\right],
\eeq
where  $D$ is the number of degrees of freedom, $x=(q,p)\in \mathbb{R}^{2D}$ is a collective notation, $\Delta H=H_{\xi_f}-H_{\xi_0}$, $x(t)$ denotes the phase space 
coordinates at time $t$ of a trajectory governed by the Hamiltonian $H_{\xi_0}$ and initial condition $x_0$, and ${\cal W}_\rho$ is the Wigner function of $\rho$:
\beq
{\cal W}_\rho(x)=\frac{1}{h^D} \int d^D\eta \;\bra{q-\eta/2} \rho \ket{q+\eta/2} e^{i p.\eta/\hbar}.
\eeq

In the quantum case, the characteristic function for a quench can be written as:
\beq
G^Q(u)=\langle e^{i u H_{\xi_f}}e^{-i u H_{\xi_0}} \rangle=\tr[e^{i u H_{\xi_f} } e^{-i  u H_{\xi_0}} \rho_0]. 
\eeq
Considering $u$ as a time-like variable,  the characteristic function can be interpreted as an evolution over a time $u\hbar$.
Then the characteristic function in the DR representation can easily be written as:
\beq
G^{DR}(u) = \int {d^{2D}x_0} \, {\cal W}_{\beta}(x_0) \, \exp\left[{\frac{i}{\hbar}\int_0^{u\hbar} \Delta H\left(x(s)\right) ds}\right], 
\eeq
where ${\cal W}_{\beta}$ is the Wigner function of the thermal state. This is equivalent to:
\beq
G^{DR}(u) = \int {d^{2D}x_0} \, \sum_m \frac{e^{-\beta E^m_{\xi_0}}}{Z^Q_{\xi_0}}  {\cal W}_{m}(x_0) \, \exp\left[{\frac{i}{\hbar}\int_0^{u\hbar} \Delta H\left(x(s)\right) ds}\right],
\eeq
where ${\cal W}_{m}$ is the Wigner function of the $m$-eigenstate of $H_{\xi_0}$.
According to the quantum ergodic conjecture (QEC), for classically chaotic systems,
\beq
{\cal W}_{m}(x)=\frac{\delta[E^m_{\xi_0}-H(x)]}{\int d^{2D}x\; \delta[E^m_{\xi_0}-H(x)]}.
\eeq
Then, using the QEC in the last equation we arrive at:
\beq
G^{SC}(u) =  \sum_m \frac{e^{-\beta E^m_{\xi_0}}}{Z^Q_{\xi_0}} \int {d^{2D}x_0} \,  \frac{\delta[E^m_{\xi_0}-H(x_0)]}{\int d^{2D}x\; \delta[E^m_{\xi_0}-H(x)]} 
 \exp\left[{\frac{i}{\hbar}\int_0^{u\hbar} \Delta H\left(x(s)\right) ds}\right].
 \label{eq:AppGsc}
\eeq
Numerically, one can approximate these integrals by sampling points with the same probability for each energy shell $R_m=\{x\in \mathbb{R}^{2D}:H(x)=E^m_{\xi_0} \}$, 
and evaluate the integral in the exponential using these initial conditions. Finally, $G^{SC}(u)$ is obtained by averaging all the integrals, each one weighted with the Boltzman factor. 
Now we will make another approximation in order to simplify the final expression that numerically is still accurate in the appropriate limit.
We will consider that $\frac{e^{-\beta E^m_{\xi_0}}}{Z}\approx\int_{E^m_{\xi_0}}^{E^{m+1}_{\xi_0}} \frac{e^{-\beta E}}{Z^C_{\xi_0}} g(E) dE$, we are also assuming that the width of the $m$th energy interval is very small, which we expect to be valid for high energy levels. 
Replacing the sum over energy levels with an integral over energies, and performing the integration (notice that $g(E)={\int d^{2D}x\; \delta(E-H(x))}$), we arrive at the following semiclassical expression for the characteristic function:
\beqa
G^{SC}(u) &\approx& \int {d^{2D}x_0} \, \sum_m \int_{E^m_{\xi_0}}^{E^{m+1}_{\xi_0}} \frac{e^{-\beta E}}{Z^C_{\xi_0}} g(E) dE \, 
 \frac{\delta[E^m_{\xi_0}-H(x_0)]}{\int d^{2D}x\; \delta[E^m_{\xi_0}-H(x)]} 
 \exp\left[{\frac{i}{\hbar}\int_0^{u\hbar} \Delta H\left(x(s)\right) ds}\right] \nonumber \\
  &\approx& \int {d^{2D}x_0} \, \int \frac{e^{-\beta E}}{Z^C_{\xi_0}} g(E) dE \,  \frac{\delta[E-H(x_0)]}{\int d^{2D}x\; \delta[E-H(x)]} 
 \exp\left[{\frac{i}{\hbar}\int_0^{u\hbar} \Delta H\left(x(s)\right) ds}\right] \nonumber \\
&\approx&  \int {d^{2D}x_0} \, \frac{e^{-\beta H_0(x_0)}}{Z^C_{\xi_0}}
 \exp\left[{\frac{i}{\hbar}\int_0^{u\hbar} \Delta H\left(x(s)\right) ds}\right].
\label{eq:GSC}
\eeqa

\subsection{Limit $\hbar\rightarrow 0$}

We will show that using this semiclassical approximation one can obtain the classical work distribution in the limit	
$\hbar\rightarrow 0$. It is easy to check that in this limit the exponential of Eq. \eqref{eq:GSC} tends to:
\beq
G_{\hbar\rightarrow 0}^{SC}(u) =  \int {d^{2D}x_0} \, \frac{e^{-\beta H_{\xi_0}(x_0)}}{Z^C_{\xi_0}} e^{i(H_{\xi_f}(x_0)- H_{\xi_0}(x_0))u}
\eeq
Now, we can include two integrals in energies:
\beq
G^{SC}_{\hbar\rightarrow 0}(u) = \int {d^{2D}x_0} \int dE_{\xi_f}\, dE_{\xi_0} \; \frac{e^{-\beta E_{\xi_0}}}{Z^C_{\xi_0}}  \delta[E_{\xi_f} - H_{\xi_f}(x_0)] \, \delta[E_{\xi_0} -  H_{\xi_0}(x_0)] \, e^{i(E_{\xi_f} -E_{\xi_0})u}
\eeq
Performing the inverse Fourier transform, multiplying and dividing by $g(E)$, and changing the order of the integrals,
it is now easy to check that this is equal to:
\beqa
P^{SC}_{\hbar\rightarrow 0}(W) &=&  \int dE_{\xi_0} \; \frac{e^{-\beta E_{\xi_0}}g(E_{\xi_0})}{Z^C_{\xi_0}} \;
\int dE_{\xi_f} \int {d^{2D}x_0}\; \frac{\delta[E_{\xi_f} - H_{\xi_f}(x_0)] \, \delta[E_{\xi_0} -  H_{\xi_0}(x_0)]}{g(E_{\xi_0})} \, \delta[W-(E_{\xi_f} -E_{\xi_0})] \nonumber \\
&=& \int dE_{\xi_0} \; \bar P_{\xi_0}^C(E_{\xi_0}) \int dE_{\xi_f} \; \bar P^C(E_{\xi_f} |E_{\xi_0})\, \delta[W-(E_{\xi_f} -E_{\xi_0})] \nonumber \\
&=& P^C(W)
\eeqa
\end{document}